\documentclass[letterpaper,10pt]{article}
\usepackage{osameet2}

\usepackage{amsmath,amssymb}

\usepackage[dvips,colorlinks=true,bookmarks=false,citecolor=blue,urlcolor=blue]{hyperref} 

\begin{document}
\bibliographystyle{unsrt}
\title{Colliding pulse mode-locked all polarization maintaining fiber gyrolaser}

\author{Hanieh Afkhamiardakani$^\dag$, Jean-Claude Diels$^\dag$}
\address{$^\dag$University of New Mexico, CHTM 1313 Goddard SE, Albuquerque, NM 87106\\
$^\ddag$University of New Mexico}
\email{jcdiels@unm.edu}

\vspace{-0.3cm}
\begin{abstract}
A bidirectional all polarization maintaining fiber laser as a highly sensitive sensor for rotation is presented. This sensor is based on
intracavity phase interferometry with application in inertial sensing,
magnetometry, displacements and nonlinear index. 
The bios beat note of this gyroscopic laser is lowered to 166 KHz implementing a symmetric configuration including two parts of Er-doped fibers. 
\end{abstract}


\section{Introduction}
Frequency combs are revolutionizing meteorology and sensing, providing absolute frequency calibration over the optical spectrum, as well as a link between optical and RF frequencies.
 Sensing is generally performed by interfering a reference frequency comb with one that has been affected by a physical quantity to be sensed.
 However, if the sensing is performed intra-cavity, and the mode-locked laser is designed with two intracavity circulating pulses,
considerably higher sensitivity than by traditional extracavity detection can be achieved.
 A passive interferometer, such as a Michelson, measures phase differences by converting them into an amplitude modulation.
 By contrast, in the Intracavity Phase Interferometer (IPI) \cite{Arissian14b}, a minute phase shift is converted into a frequency.
  This frequency is that of a beat note obtained by interfering the two frequency combs issued from the same laser.
 The signal to noise ratio of this beat note measurement is considerably higher than with extracavity sensing. This is due to the fact that the noise of the two interfering combs is correlated.
\subsection{Implementing IPI in laser gyroscops}
Bidirectional mode-locked lasers are very promising for implementation of highly sensitive intracavity phase interferometry for high precision sensing.
Examples of  applications are detection of small displacement, linear and nonlinear refractive indices, magnetic field,
 scattering, rotation, and acceleration.
Rotation sensors or laser gyroscopes are of great interest to develop modern navigation systems.
Although gyroscopic response was successfully realized in discrete element bidirectional mode-locked dye lasers \cite{Arissian14b, Dennis91b,Hendrie16}, it is impractical in commercial systems.
Fiber lasers are the most promising laser media to implement IPI owing to the possibility of producing ultrashort pulses with a compact and low-cost design \cite{Afkhamiardakani18B,Hendrie18, Kieu08, Krylov17}.
However, the nonlinear effects distributed over long distances in fiber lasers poses a challenge to the implementation of IPI in fiber lasers.
 Unless extreme care is taken to design a rigorously symmetric structure and defining a very thin layer of saturable absorber, the crossing point of two
  circulating pulses in a passively mode-locked device will not be at the location of the
 saturable absorber to achieve colliding pulse mode-locking.
The operation of single mode fiber lasers is also very sensitive to external perturbations such as motion,
 vibration, thermal change, and any other parameter which can contribute to polarization change through
 motion or stresses, which limits the stability of mode-locking in single mode fiber lasers.

In this work we overcome these issues by constructing an all-polarization maintaining bidirectional mode-locked fiber laser with two portions of Er-doped fibers
 pumped through WDMs to eliminate all the asymmetries in the cavity. A fine control of the bias beat note is obtained by tuning the pump powers of the two gain
  sections, in order to minimize the difference between the nonlinear phase accumulated in each direction between absorber and output coupler.
  \subsection{Formulation of gyroscopic effect}
The beat note frequency is obtained by interfering two trains of fs pulses issued from a laser.
 Assuming P being the cavity perimeter
  and $\nu$ the light frequency,
 $\Delta P=P\Delta\nu/\nu$
   corresponds to an optical path difference between opposite directions. On the other hand, in a rotating system with angular velocity of
    $\omega_{r}$, the pulse circulating in
     the same direction travels $\Delta P$
      more than the one in opposite direction which is given by
      $\Delta P=2R\omega_{r}P/c$
      where R is the radius of laser gyro and c is the speed of light. Therefore, the gyroscopic response is calculated as
       $\Delta\nu=2R \omega_{r}/\lambda$.
\section{Bidirectional mode-locked fiber laser}

As mentioned in previous section, there are numerous challenges associated with the implementation of IPI
in fiber lasers. Since a typical fiber laser operates with large gain and losses,
the pulses circulating in opposite direction in a ring laser traverse the optical components
(gain, output coupling, saturable absorption) in a
different order.  The resulting asymmetry of the counter-circulating pulses makes a huge bias beat note. Therefore,
as a first step to reduce the bias of the system, we designed a very symmetric cavity.

\subsection{Experimental setup}
\begin{figure}[h!]
  \centering
  \includegraphics[width=0.8\linewidth]{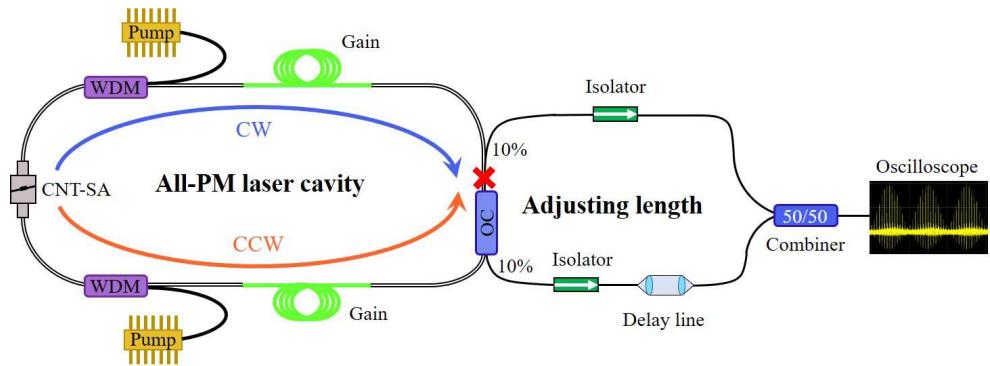}
\caption{\small Experimental setup for colliding pulse mode-locked all polarization maintaining fiber gyrolaser. WDM: wavelength division multiplexer, OC: output coupler, CNT-SA: carbon nanotube saturable absorber.}
\label{fig:setup}
\end{figure}
Fig.~\ref{fig:setup} shows a schematic configuration of a bidirectional fiber laser which is used as a gyrolaser to sense rotation. This laser includes a total length of
102 cm of Er-doped fiber  gain medium  distributed
 in two different locations of the laser, symmetrically to the saturable absorber. These are pumped by laser diodes at 980 nm via polarizaton maintaining (PM) wavelength division
multiplexers (WDM). 
The saturable absorber (SA) is a thin layer of carbon nanotubes (CNT) squeezed between two angled-cut fiber ferrules which establishes the crossing point of the counter-circulating pulses.
Tapered fiber covered with CNTs is another option to mode-lock fiber lasers \cite{Afkhamiardakani16}
but it fails to stabilize the crossing point of the two pulses in bidirectional lasers.

Using WDMs in reflection helps to protect SA from overheating by filtering out the extra power from pump lasers as carbon nanotubes
are easily burned due to direct high intensity irradiation which also makes the mode-locking unstable \cite{Afkhamiardakani18A}.
A 2 by 2 PM output coupler extracts 10\% of the light from the fiber laser cavity on each direction.
The output pulse trains from clock-wise (CW) and counter clock-wise (CCW) directions combine through a 50/50 combiner.
A home-made delay line is placed in the shorter direction to compensate the length difference in micrometer.
The intensity of the counter-circulating pulses can be adjusted through  the pump powers in order to minimize the bias beat note of the gyrolaser.

\subsection{Results}
\begin{figure}[h!]
  \centering
  \includegraphics[width=0.8\linewidth]{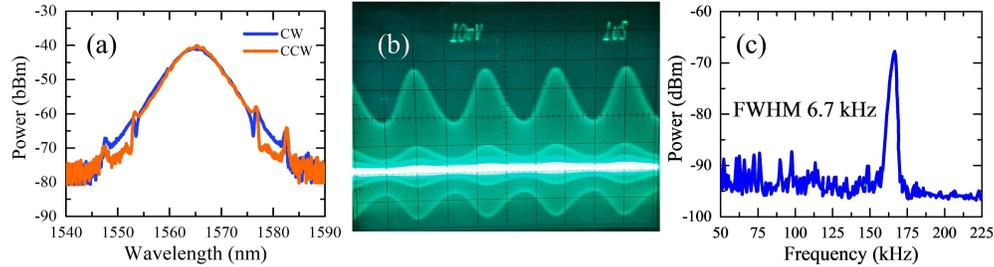}
\caption{\small a) Optical spectra of counter-circulating pulses of the laser gyro b) creation of sinusoidal pattern on the oscilloscope due to superposition of the pulse trains corresponding to slightly shifted frequency combs c) Radio frequency spectrum with resolution bandwidth of 3 kHz}
\label{fig:result}
\end{figure}

Fig. \ref{fig:result}(a) shows the soliton spectra of the counter-circulating pulses in the mode-locked laser.
It should be mentioned that the central wavelengths of pulses are almost equal within the 0.5 nm resolution band of our spectrum analyzer. The output average powers are 484 and 680 $\mu$W for CW and CCW directions, respectively.
As shown in Fig.~\ref{fig:result}(b), superposition of the pulse trains corresponding to slightly shifted frequency combs
 creates a sinusoidal pattern on the oscilloscope. The radio frequency
 spectrum which is basically the Fourier transform of Fig.~\ref{fig:result}(b) gives the beat note frequency plotted in Fig.~\ref{fig:result}(c).
The beat note frequency is a measure of the shift between two frequency combs in opposite directions. The circulating pulses have two
 crossing points located at the SA,  and just before the output coupler shown by cross sign in Fig.~\ref{fig:setup}. In this case the bias of the laser
  falls below its dead band which is mostly arisen from phase coupling at the absorber. Adding more PM
  fiber in cw direction breaks the symmetry. The lowest measurable beat note is 166 kHz.
  It is essential to minimize the dead band of the laser created by the large scattering at
  the position of the SA. The potential solution is to force one of the intracavity pulses to
   propagate along the fast axis of the PM fiber and the other along the slow axis. In this case,
   even if the light scatters from SA, its polarization is orthogonal and the least phase
   coupling happens between cross polarized, counter-propagating pulses.

%\bibliography{c:/bib/ad,c:/bib/en,c:/bib/oz,c:/bib/ref-Hanieh}

\end{document}